\documentclass[12pt]{article}

\usepackage{newtxtext,newtxmath}

\usepackage{graphicx}
\usepackage{siunitx}
\usepackage{booktabs}

\usepackage[letterpaper,margin=1in]{geometry}

\renewenvironment{abstract}
	{\quotation}
	{\endquotation}

\date{}

\makeatletter
\renewcommand{\fnum@figure}{\textbf{Figure \thefigure}}
\renewcommand{\fnum@table}{\textbf{Table \thetable}}
\makeatother

\usepackage{scicite}

\usepackage{url}

\def\scititle{Nucleon Strong-Interaction Size From Charmonium Photoproduction}

\title{\bfseries \boldmath \scititle}

\author{
    Xiangdong~Ji$^{1,2}$, 
    Sylvester~Joosten$^{3}$, 
    Zein-Eddine~Meziani$^{3\ast}$,
    Dimitra~A.~Pefkou$^{4,5,6}$\and
    \small$^1$Tsung-Dao Lee Institute and School of Physics and Astronomy,\and
    \small Shanghai Jiao Tong University, Shanghai 200240, China\and
    \small$^2$Department of Physics, University of Maryland, College Park, MD 20742, USA\and
    \small$^3$Physics Division, Argonne National Laboratory, Lemont, IL 60439\and
    \small$^4$Department of Physics, University of California, Berkeley, CA 94720, U.S.A\and
    \small$^5$Nuclear Science Division, Lawrence Berkeley National Laboratory, Berkeley, CA 94720, USA\and
    \small$^6$National Energy Research Scientific Computing Center, \\ \small Lawrence Berkeley National Laboratory, Berkeley, CA 94720, USA\and
	\small$^\ast$Corresponding author. Email: zmeziani@anl.gov }

\begin{document} 

\maketitle

\begin{abstract} \bfseries \boldmath
 
Protons and neutrons differ dramatically in their electromagnetic sizes, but are bound by the strong (color) force and have nearly identical masses, like two twin states of the same “nucleon”. To reveal the overall spatial extent, it must be examined through the strong interaction rather than electromagnetism. The $J/\psi$ particle, a compact color-charge dipole of a charm quark and antiquark, provides an ideal probe. Recent high-precision measurements of near-threshold $J/\psi$ production at Jefferson Lab offer a more precise glimpse of the proton’s gluonic spatial structure. Combining these data with recent quark-sector results, we extract the distribution of the color field underlying mass generation in the nucleon and find it extends beyond both charge and mass distributions, revealing the nucleon’s overall strong-interaction spatial extent of $0.94\pm 0.09$ fm.

\end{abstract}

\noindent
What is the size of the proton? This seemingly simple question is often answered in terms of its electric charge radius, measured with increasing precision over decades through elastic electron scattering and other electromagnetic probes \cite{Karr:2020wgh,Gao:2021sml}. Yet the proton is more than just a charged particle—it is a strongly interacting system of quarks and gluons, whose dynamics govern its internal structure, with electrically neutral gluons playing a key role in its mass, spin, and energy distribution.
 
The proton’s nearly identical partner, the neutron, shares almost the same mass and strong-force properties, and should therefore have a similar overall strong-interaction size. However, its charge radius is nearly zero due to charge cancellation between up and down quarks, suggesting strongly that it is not an ideal measure of the nucleon overall spatial extent. A comparable situation arises in heavy nuclei such as $^{208}$Pb, where the neutron distribution extends beyond the proton one, forming a “neutron skin”, implying that both protons and neutrons must be considered when defining the overall spatial extent of \(^{208}\)Pb.

Thus, the answer to the opening question must come from a strong-interaction “microscope” capable of probing inside nucleons and mapping the color, or gluonic, field — the messenger of the strong force. An ideal probe is a beam of particles composed of a heavy quark–antiquark pair, known as quarkonium, which forms a compact color dipole sensitive to gluon distributions \cite{Kaidalov:1992hd}. However, heavy quarkonia are unstable and can be produced during high-energy interactions involving photons or electrons. Because such collisions often destroy the target nucleon, experiments must be finely tuned to just above the energy needed to create quarkonia, a regime known as near-threshold production. These processes are rare and require intense beams. The upgraded 12-GeV Jefferson Lab has recently made it possible, producing high-precision data on near-threshold $J/\psi$ photoproduction off the proton \cite{Duran:2022xag,Joosten:2025,Ali:2019lzf,GlueX:2023pev,CLAS:2026lls}, 
enabling an extraction of the nucleon’s strong-interaction size from the world \(J/\psi\) photoproduction data within the holographic QCD framework.

To achieve this, two crucial intermediate steps are required: First, we must perform a global analysis of the available data with reasonable control over modeling the probe. Second, we must connect the extracted quantities to the relevant nucleon strong-interaction size. 

Near Threshold \(J/\psi\) photoproduction off the proton has been proposed as a sensitive probe
of the gluonic ``stress-energy tensor'' (SET) inside the nucleon~\cite{Kharzeev:2021qkd,Guo:2021ibg,Mamo:2019mka,Hatta:2019lxo,Mamo:2022eui,Ji:2020bby,Sun:2021gmi}. In particular, within the holographic QCD framework adopted in this work, the near-threshold production amplitude is related to the gluonic SET form factors through the exchange of the corresponding gluonic modes. We note, however, that this interpretation has some level of framework dependence. Alternative production mechanisms, including contributions from open-charm intermediate states, have been proposed~\cite{Du:2020bqj,JointPhysicsAnalysisCenter:2023qgg}, which may complicate a model-independent identification of the measured photoproduction amplitude with the gluonic SET.

The spatial distributions encoded in the gluonic SET are characterized by three ``form factors", \(A_g(t)\), \(B_g(t)\), and \(C_g\) (or, equivalently \(D_g(t)\))~\cite{Pagels:1966zza,Kobzarev:1962wt,Ji:1996ek}, where \( t \) is the squared momentum transfer in the scattering and a Fourier conjugate of the position variable. \(B_g\) is related to the spin structure of the nucleon~\cite{Ji:1996ek} and is generally neglected as it is found to be small~\cite{Shanahan:2018pib,Pefkou:2021fni,Hackett:2023rif}. The form factors \(A_g\) and \(C_g\) (or, equivalently \(D_g\) ) provide the information of the color field energy densities and stress (or more precisely the momentum current) distributions~\cite{Polyakov:2002yz,Ji:2021mtz,Ji:2025gsq}. Therefore, our first task is to use the world available data to make a globally consistent extraction of \(A_g\) and \(C_g\) (or equivalently \( D_g\)) form factors. To subsequently connect them to an appropriate strong-interaction size, we first need to understand a very important phenomenon in QCD: the ``trace anomaly''.

The SET $T^{\mu\nu}$, where $\mu,\nu=0,1,2,3$ are space-time indices, contains crucial mechanical information about a system. As in electromagnetism, the quantum chromodynamics (QCD) color-field SET  $T^{\mu\nu}_g$ is made of bilinear chromo-electric ($\vec{E}^a$) and chromo-magnetic fields ($\vec{B}^a$) encoded in the gluon field strength $G^{a\mu\nu}$, with $a=1,..,8$. It is traceless in classical theory, i.e., $T^\mu_{g~\mu}=0$ with $\mu$ summed over, due to the fact that gluons, just like photons, are massless. Thus, strong and electromagnetic interactions do not seem to have their own mass or distance scale! While this is true in electromagnetism where the scale is entirely set by the electron mass, we know from phenomenology that in strong interactions the scale is not simply decided by the quark masses.
 
What comes to the rescue is a remarkable feature of quantum field theory known as an anomaly~\cite{Adler:1969gk,Bell:1969ts}. Unlike in classical theory, the quantum fluctuations at extremely small distances, i.e., at scales much smaller than the size of the nucleon, induce a trace in the gluonic SET~\cite{Collins:1976yq,Nielsen:1977sy},  
 \begin{equation} 
   T^\alpha_{g~\alpha } = \frac{\beta(g)}{2g} G^{a\mu\nu} G^a_{\mu\nu}  \equiv G^2
 \label{eq:trace}
 \end{equation}
where \(\beta(g)\) is the so-called QCD beta function controlling the scale dependence of strong interaction coupling \(g\), resulting from the ultra-violet quantum fluctuations. This trace term, or scalar color field $G^2$, generates the scale of strong interaction. Therefore, nothing seems more appropriate than to measure the size of the nucleon through the color field $G^2$ or, equivalently, through the distribution of the trace-anomaly scale inside it. In the much-studied bag models for the nucleon in the 1970's, the $G^2$ distribution is modeled as a ``bag'' in which the quarks are confined~\cite{Bogolubov:1968zk,Chodos:1974je,DeGrand:1975cf,Thomas:1981vc}. 

It comes as no surprise that the scalar (color) field plays a central role in determining the fundamental properties of the nucleon. A similar mechanism operates in the electroweak theory in which the scalar Higgs field, with non-vanishing vacuum condensation,
generates the energy scale for electroweak interactions that gives the masses of leptons and other elementary particles~\cite{Higgs:1964pj,Englert:1964et,Guralnik:1964eu}. In terms of scale setting, the color field $G^2$ plays a similar role in the strong interaction sector as the Higgs in the electroweak sector. For instance, the light hadron mass $M_H$ is a combined SET trace effect from the gluonic $T^\mu_{g~\mu}$ and the quark mass term from $T^\mu_{q~\mu}=\sum_q (1+\gamma_m) m_q \bar{q} q $, where \(m_q\) is the quark mass,

 \begin{equation}
 M_H\sim \langle H|T^\mu_{g~\mu} + T^\mu_{q~\mu} |H\rangle   \ ,  
 \end{equation}
 and $|H\rangle$ is the hadron state. For the nucleon, the weak-interaction contribution is less than 10\% because of the small up, down (and strange) quark couplings to the Higgs, and therefore $M_H\sim \langle H|G^2|H\rangle$, as in Higgs theory\footnote{While the color field sets the fundamental scale, the actual mass of baryons including the nucleon, has contributions from quark and gluons kinetic motion, with direct contributions from the color field amounting to only about a quarter of baryon mass~\cite{Ji:1994av,Ji:2021mtz,Ji:2021pys}.}. 

A direct measurement of the scalar color field $G^2$ distribution in the nucleon, or equivalently of the corresponding form factor denoted as $G^s(t)$, is an open and challenging problem. This is because its impact on high-energy scattering processes is sub-leading and highly suppressed, although attempts at finding a path to access it directly have been discussed in literature~\cite{Hatta:2023fqc,Hatta:2018sqd}. Fortunately, there is an indirect approach provided by the law of energy-momentum conservation: Even though the quantum fluctuation generates an anomaly term in the trace of the total SET, which includes both quark and gluon contributions as seen in Eq.(2), the entire tensor still satisfies $\partial_\mu T^{\mu\nu}=0$. That means that  $G^s(t)$ is rigorously related to the three SET form factors (FFs), $A(t)$, $B(t)$ and $C(t)$~\cite{Goeke:2007fp,Ji:2021mtz}. Here $A\equiv A_{q+g}$, $B\equiv B_{q+g}$, and $C\equiv C_{q+g}$ refer to the total SET FFs, which include both quark and gluon contributions, though until recently very little was known about the latter piece (see Ref.~\cite{Burkert:2023wzr} for a recent review). From these FFs, one can also define the proton's total mass form factor~\cite{Goeke:2007fp,Ji:2021mtz}, $G^m(t)$, which is related to its energy density and corresponding radius (mass radius), just like $G^s(t)$ is related to its scalar color field density and radius. Refer to the Materials and
Methods~\cite{SM} for more details.

We now have a clear road-map towards determining the strong interaction size of the nucleon. We first determine two of the gluon SET FFs, \( A_g(t) \) and \( C_g(t) \), from the modern near threshold \( J/\psi \) photoproduction off the proton data~\cite{Duran:2022xag,Joosten:2025,Ali:2019lzf,GlueX:2023pev, CLAS:2026lls}. Combining this critical 
information with their quark counterparts from a recent global Bayesian extraction~\cite{Guo:2025jiz}, we can evaluate the total scalar color field and mass form factors, and define their density distributions and the strong-interaction and mass radii in the Breit frame, which has been a standard practice for the charge radii in the field~\cite{Sachs:1962zzc,Polyakov:2002yz}.

We adopt an effective theory framework for the scattering that allows extractions of the gluon SET FFs in a broader kinematic region.
In order to perform our global fit of the differential cross sections in the full energy range close to threshold 
and determine \( A_g(t)\) and \( C_g(t)\), we choose to use the holographic QCD model~\cite{Mamo:2022eui}, where \(A_g(t)\)  and  \(C_g(t)\) are modeled as 
a dipole and tripole function respectively, 
dictated by the perturbative QCD analysis of the form factors behavior at large momentum transfer~\cite{Tong:2022zax,Sun:2021gmi,Tanaka:2018wea}. We assume that \( B_g(t) = 0 \), consistent with lattice QCD calculations~\cite{Shanahan:2018pib,Pefkou:2021fni,Hackett:2023rif}, and excluded in the holographic QCD model~\cite{Mamo:2022eui}.

\begin{figure} 
	\centering
	\includegraphics[width=0.8\textwidth]{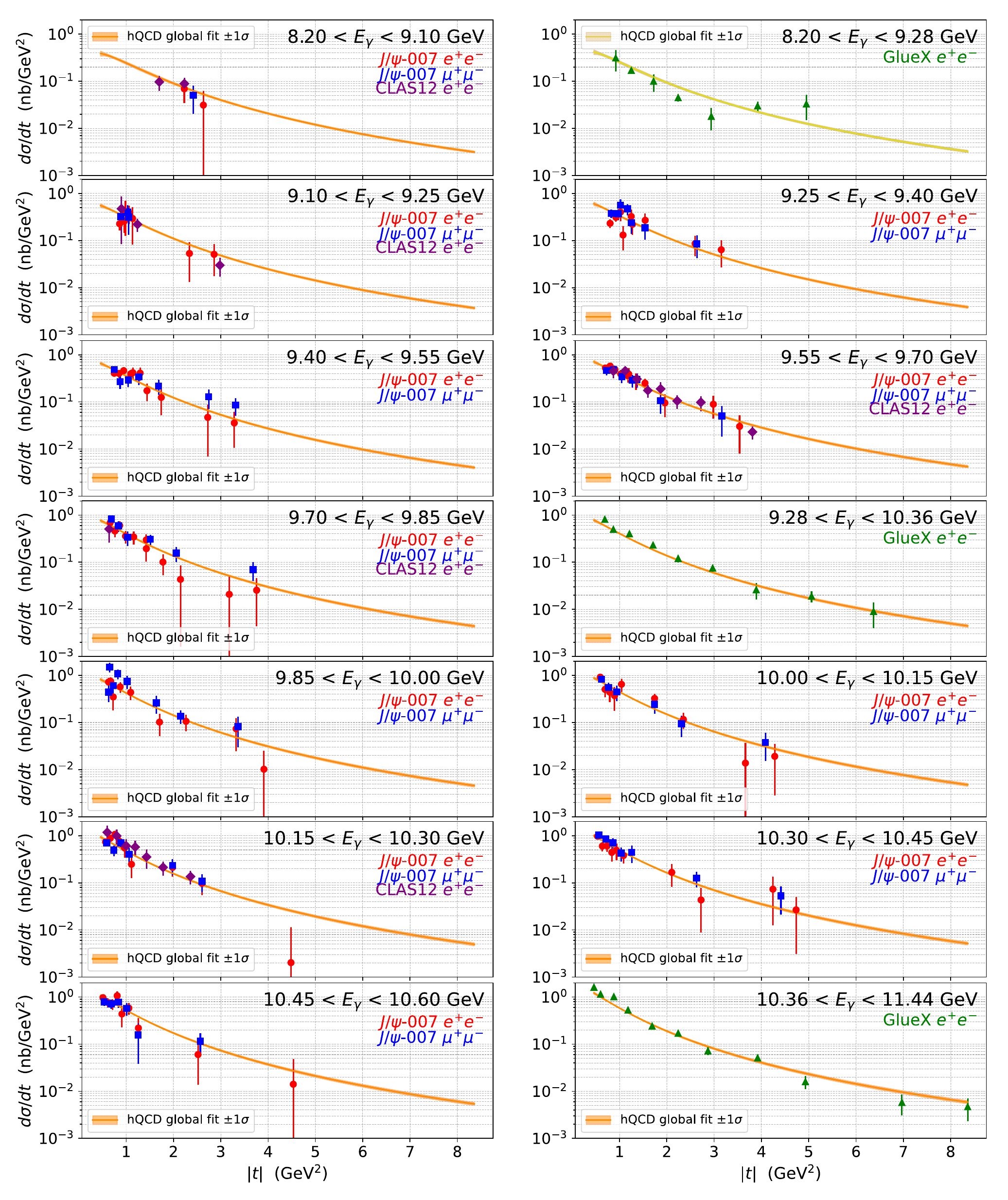} 

  \caption{{\bf Differential cross sections and global holographic QCD fit.} \\
    World data on near-threshold \( J/\psi \) photoproduction off the proton, shown as a function of momentum transfer \( |t| \) in bins of photon energy \( E_\gamma \). The orange curves and shaded bands represent the global fit without the GlueX lowest photon energy data (top right panel) and its uncertainty using the holographic QCD model~\cite{Mamo:2022eui} with a dipole–tripole form for the SET FFs. The light-green curve on the top right panel is the result of the fit compared to the excluded lowest GlueX photon energy data. The fit describes the data well across the full range; excluding the lowest-energy GlueX bin (top right) improves the fit quality from \( \chi^2/\text{n.d.f.} \) from 1.070 (see Fig.~S5) to 1.005 (see Fig.~S6). The results of this fit compared to other data sets including the GlueX lowest energy are shown in~\cite{SM}.}
 \label{fig:tdistribution}
 \end{figure}

Fig.~\ref{fig:tdistribution} shows all the available differential cross section data compared to our fit used to extract the parameters of the \( A_g \) and \( C_g \) gluonic SET FFs. Previous attempts of fitting
the relevant experimental data by theorists are all limited
by either the applicable ranges of scattering models or partial experimental data~\cite{Guo:2021ibg,Kharzeev:2021qkd,Mamo:2021krl,Guo:2023pqw}. 
More details on the datasets, cross section model,  theoretical considerations, and fitting procedure can be found in~\cite{SM}. 
To obtain the total scalar color field and mass distributions, we combine our global experimental determination for the gluon SET FFs with a recent Bayesian inference of the quark SET FFs~\cite{Guo:2025jiz} from recent experimental and lattice QCD data. We note that the results of the fits in that study are consistent with earlier extractions that used different approaches and relied solely on experimental data~\cite{Pasquini:2014vua,Dupre:2017hfs,Burkert:2018bqq}.

\begin{figure}
\centering
    \includegraphics[width=0.9\textwidth]{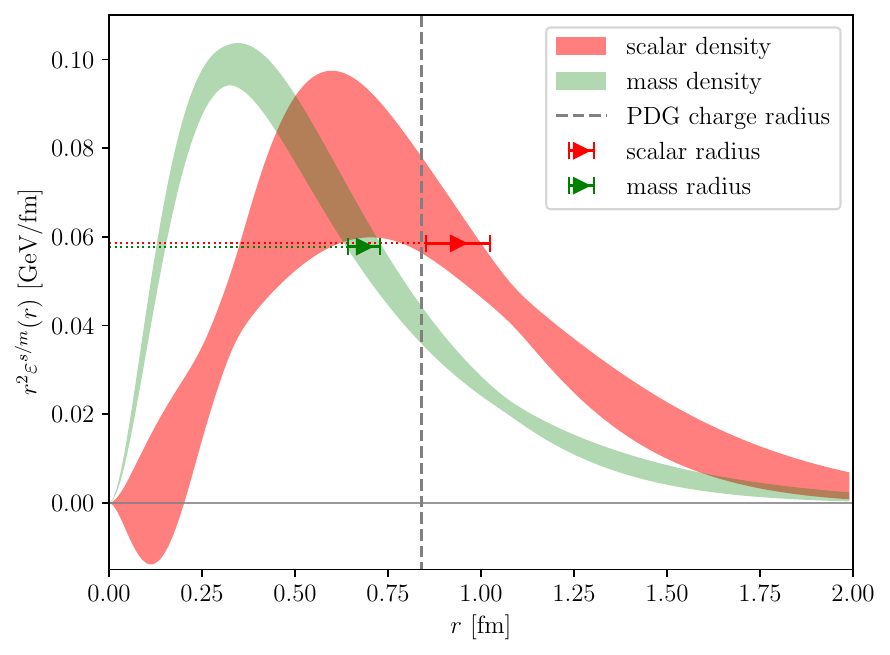}
\caption{{\bf Densities and radii of the proton.} \\
A comparison of different measures of the proton size; the scalar and mass radii, with their corresponding scalar ($\varepsilon^s(r)$) and mass ($\varepsilon^m(r)$) density profiles, are obtained by combining gluon contributions extracted from a global analyses near-threshold \( J/\psi \) photoproduction and with quark contributions from a Bayesian GFF analysis~\cite{Guo:2025jiz}. These are compared against the canonical proton size measure, its electric charge radius ($0.8409\pm 0.0004$~fm), using the latest PDG average~\cite{ParticleDataGroup:2024cfk}. The scalar strong-interaction radius ($0.94 \pm 0.09$~fm) is the largest of the three, reflecting the realistic spatial extent of the nucleon.}
\label{fig:densities_main}
\end{figure}

In Fig.~\ref{fig:densities_main}, we present the proton's 
scalar (red) and mass (green) energy density profiles in the Breit frame. The integral of the scalar density is responsible for the anomalous energy in the proton mass decomposition~\cite{Ji:1994av,Ji:2021pys}. The strong-interaction (dominated by the scalar color field) and the mass radii in the Breit Frame are defined as\footnote{Note that the former is different than the radius of the quark scalar density $\bar{q}q$ obtained by the $\sigma_{\pi N}$ form factor~\cite{Hoferichter:2012wf,Hoferichter:2016nvd,Hoferichter:2023ptl}, although the latter is a part of the contribution to the strong radius through small quark masses as in Eq.(2)}:
\begin{align} \label{eq:radii}
\langle r_s^2 \rangle &=  
 \left. 6 \frac{dA(t) }{dt} \right|_{t=0} -
\frac{36}{2M^2}  C(0),\nonumber \\
\langle r_m^2 \rangle &=  
\left. 6 \frac{dA(t)}{dt}\right|_{t=0} -
\frac{12}{2M^2}  C(0).
\end{align}
Our findings indicate that the
strong-interaction size is close to \(\sim\)1 fm, larger than the charge and mass radii, and primarily governed by the scalar color field. Although this has been expected from bag model phenomenology in which the scalar potential (or bag constant) is present to confine the motion of quarks~\cite{Bogolubov:1968zk,Chodos:1974je,Thomas:1981vc}, this allows us to infer the corresponding physical spatial distribution from the experimental data within the theoretical framework employed here.

Of course, the scalar color field is present in the strong-interaction vacuum already as a condensate~\cite{Shifman:1978bx,Shifman:1978by}, again similar to the Higgs field. However, what we present here is its response to the presence of valence quarks, forming an attractive potential (the bag) to confine the latter. Much like the \(^{208} \)Pb neutron ``skin", the nucleon exhibits a scalar color-field ``skin"---quantified by the difference \(\sqrt{\langle r_{\rm scalar} ^2\rangle} - \sqrt{\langle r_{\rm charge}^2\rangle}\) between the strong-interaction and charge radii---underscoring the essential role of gluons in defining the nucleon's size and internal dynamics.

Although our finding of $\langle r_s^2 \rangle=(0.94 \pm 0.09{\rm fm})^2$ is still limited by the statistical and model uncertainties, it represents the first comprehensive study of the
nucleon's strong interaction radius from a
proper experimental probe. The radius appears stable, even when excluding some of the datasets, as discussed in~\cite{SM}. The result has comparable accuracy with the lattice QCD and other theoretical analyses~\cite{Hackett:2023rif,Wang:2024lrm,Cao:2024zlf}, which seem to place the strong-interaction radius outside the charge radius, which in turn is larger than the mass radius. 
Just as the precision of the proton charge-radius determination has improved dramatically since Hofstadter’s pioneering measurements, we anticipate that future high-precision experimental probes of color interactions at Jefferson Lab with the SoLID detector~\cite{JeffersonLabSoLID:2022iod} and at the EIC with the ePIC detector~\cite{Gryniuk:2020mlh} will improve the precision of the nucleon’s strong-interaction radius determination by up to an order of magnitude.

\newpage


\clearpage 

%
\bibliography{science-advances-references} 
\bibliographystyle{sciencemag}

%
%
%
%
%
%


\section*{Acknowledgments}
We thank K. Mamo and I. Zahed for valuable discussions and support on the holographic QCD approach. We thank D. Hackett and P. Shanahan for collaborating on the generation and analysis of lattice results used in this work. We thank the authors of refs.~\cite{Guo:2025jiz} and~\cite{Wang:2024lrm} for sharing their data.
\paragraph*{Funding:}
X.J.'s work is supported partially by Maryland Center for Fundamental Physics. S.J. and Z.-E.M.'s work is supported in part by the US Department of Energy Office of Science, Office of Nuclear Physics under contracts numbers DE-AC02-06CH11357. D.A.P.'s work is supported from the Office of Nuclear Physics, Department of Energy, under contract DE-SC0004658. 

\paragraph*{Author contributions:}
The authors have contributed equally to this article 

\paragraph*{Competing interests:}
The authors declare no competing interests.

\paragraph*{Data and materials availability:}
The data used in this work are all publicly available from their respective cited primary publications.





\end{document}